\documentclass[journal=nalefd,manuscript=letter]{achemso}

\usepackage[version=3]{mhchem} 
\usepackage{color}

\author{A.~A.~Kozikov}
\email{akozikov@phys.ethz.ch}
\affiliation{Solid State Physics Laboratory, ETH Z\"{u}rich, CH-8093 Z\"{u}rich, Switzerland}
\author{R.~Steinacher}
\affiliation{Solid State Physics Laboratory, ETH Z\"{u}rich, CH-8093 Z\"{u}rich, Switzerland}
\author{C.~R\"{o}ssler}
\affiliation{Solid State Physics Laboratory, ETH Z\"{u}rich, CH-8093 Z\"{u}rich, Switzerland}
\author{T.~Ihn}
\affiliation{Solid State Physics Laboratory, ETH Z\"{u}rich, CH-8093 Z\"{u}rich, Switzerland}
\author{K.~Ensslin}
\affiliation{Solid State Physics Laboratory, ETH Z\"{u}rich, CH-8093 Z\"{u}rich, Switzerland}
\author{C.~Reichl}
\affiliation{Solid State Physics Laboratory, ETH Z\"{u}rich, CH-8093 Z\"{u}rich, Switzerland}
\author{W.~Wegscheider}
\affiliation{Solid State Physics Laboratory, ETH Z\"{u}rich, CH-8093 Z\"{u}rich, Switzerland}

\title{Mode specific backscattering in a quantum point contact}


\begin{document}
\begin{abstract}
We demonstrate a scanning gate grid measurement technique consisting in measuring the conductance of a quantum point contact (QPC) as a function of gate voltage at each tip position. Unlike conventional scanning gate experiments, it allows investigating QPC conductance plateaus affected by the tip at these positions. We compensate the capacitive coupling of the tip to the QPC and discover that interference fringes coexist with distorted QPC plateaus. We spatially resolve the mode structure for each plateau.

\end{abstract}

\section{Introduction}

Scanning gate microscopy \cite{Eriksson, SellierReview, FerryReview} (SGM) is used to investigate the conductance of a nanostructure by modifying its electrostatic potential \cite{Crook2002, Woodside, Pioda2004, Fallahi, Hackens, Bleszynski, Schnez, KozikovStadium, KozikovAB, Brun} (gating effect) or by locally changing the electron density of a two-dimensional system \cite{Crook2000a, Crook2000b, TopinkaSci, TopinkaNat, Aoki, Konig} resulting in backscattering in case of a strong tip potential.
Both effects usually coexist (see e.g. Refs.\cite{JuraPRB, KozikovBranches} for detailed discussions) and appear due to a Lorentzian potential \cite{Kicin, Pioda, Gildemeister, Steinacher2014a} induced by the electrically biased tip of the scanning force microscope. In top gate defined quantum point contacts (QPCs) \cite{Wees, Wharam} such backscattering experiments allow studying quantum effects, e.g. the interference of backscattered electrons \cite{TopinkaSci, TopinkaNat, LeroyAPL, LeroyPRL, JuraNatPhys, JuraPRB, Paradiso, KozikovBranches} seen as small variations of the conductance in the form of interference fringes decorating branches. The gating effect (modification of the QPC potential via capacitive coupling of the tip to the QPC) significantly decreases the conductance when scanning the tip close to the QPC. Since gating is the stronger effect, quantum interference effects can be masked by gating \cite{KozikovBranches}. It is therefore important to compensate the gating effect of the QPC to obtain more precise information about the quantum effects.

One way to compensate the gating effect is to scan the surface twice \cite{JuraPRB, KozikovBranches} (the two-pass technique used in previous experiments): (i) at a large tip-surface separation, when the tip does not deplete the 2DEG, and (ii) at a small separation when it does. In both passes the capacitive coupling of the tip to the QPC (gating) is approximately the same, but backscattering is present only in pass (ii). Therefore, by recording the gate voltage necessary to keep the conductance constant in pass (i) and using this know-how in pass (ii) by simultaneously measuring the conductance allows compensating the gating effect. However, this technique may not be reliable, because often gating is slightly different for the two tip-surface separations (the difference becomes larger as the tip moves closer to the QPC), which can lead to undercompensation. It depends on the parameters of the gate voltage feedback that keeps the conductance constant in pass (i), which limits the scanning speed especially very close to the QPC where the tip potential is steep at the position of the constriction and therefore the conductance changes rapidly. Scanning the surface twice and measuring only at a single value of the conductance set in pass (i) significantly increases the time for the detailed investigation of physical phenomena compared to the time needed for single-pass scans.

In this work we present a scanning gate grid technique and demonstrate its advantages for the example of a scanning gate measurement close to a QPC fabricated in a high-mobility GaAs/AlGaAs heterostructure. The method consists in measuring the conductance of a QPC as a function of top gate voltage at each tip position. This kind of technique is well known in scanning probe microscopy. For example, in scanning tunnelling microscopy it is well-established to measure the differential conductance at each tip position (see e.g. Refs.\cite{ChenSTM, WiesendangerSTM}). In previous SGM experiments, conductance--tip voltage traces were recorded at several tip positions to study the quantized conductance in a quantum wire \cite{CrookGrid}, defects in carbon nanotubes \cite{Hunt}, the conductance in the quantum Hall regime in a QPC \cite{MartinsQHE}. We exploit the grid technique using a large density of points in a three-dimensional parameter space.

Scanning gate grid measurements allow fully compensating the gating effect at any tip-QPC distances. The only limitation to scan close to the QPC is the size of the tip-depleted region \cite{KozikovBranches}. A single grid measurement contains scanning gate data for many QPC transmissions. The manipulation of data, e.g. subtracting the conductance at different gate voltages, is more reliable compared to that in standard SGM measurements. It is carried out at each tip position. Since it takes much less time to record values of the conductance at different gate voltages at a specific tip position using the grid than the standard technique, sample drifts will affect the results much less. In addition, measuring the gate voltage dependent conductance at each tip position gives information about the effect of the tip on these traces, which is inaccessible in the standard measurements.
Scanning as close to the QPC as possible we resolve its mode structure seen as a widening lobe pattern with the number of lobes equal to the number of QPC modes and maxima of $|\Psi|^2$ inside the point contact up to five modes. We observe distorted conductance plateaus caused by the tip and discuss their origin. We study the fringe spacing for the QPC modes and distorted plateaus. Results of the SGM grid measurements are compared to previous scanning gate experiments. Grid measurements are thus a powerful tool which can give more information about the studied system compared to that in standard scanning gate experiments.

\section{Results and discussion}

A two-dimensional electron gas 120 nm deep below the surface is defined in a high-quality GaAs/AlGaAs heterostructure. At a density of $n = 1.4\times10^{11}$ cm$^{-2}$ electrons have a mobility of $9.6\times10^6$ cm$^2$/Vs measured at 300 mK. Their elastic mean free path and the Fermi wavelength are $l_\mathrm{p}=60~\mu$m and $\lambda_\mathrm{F}=66$ nm, respectively.

\begin{figure}[H]
\begin{center}
\includegraphics[width=\textwidth]{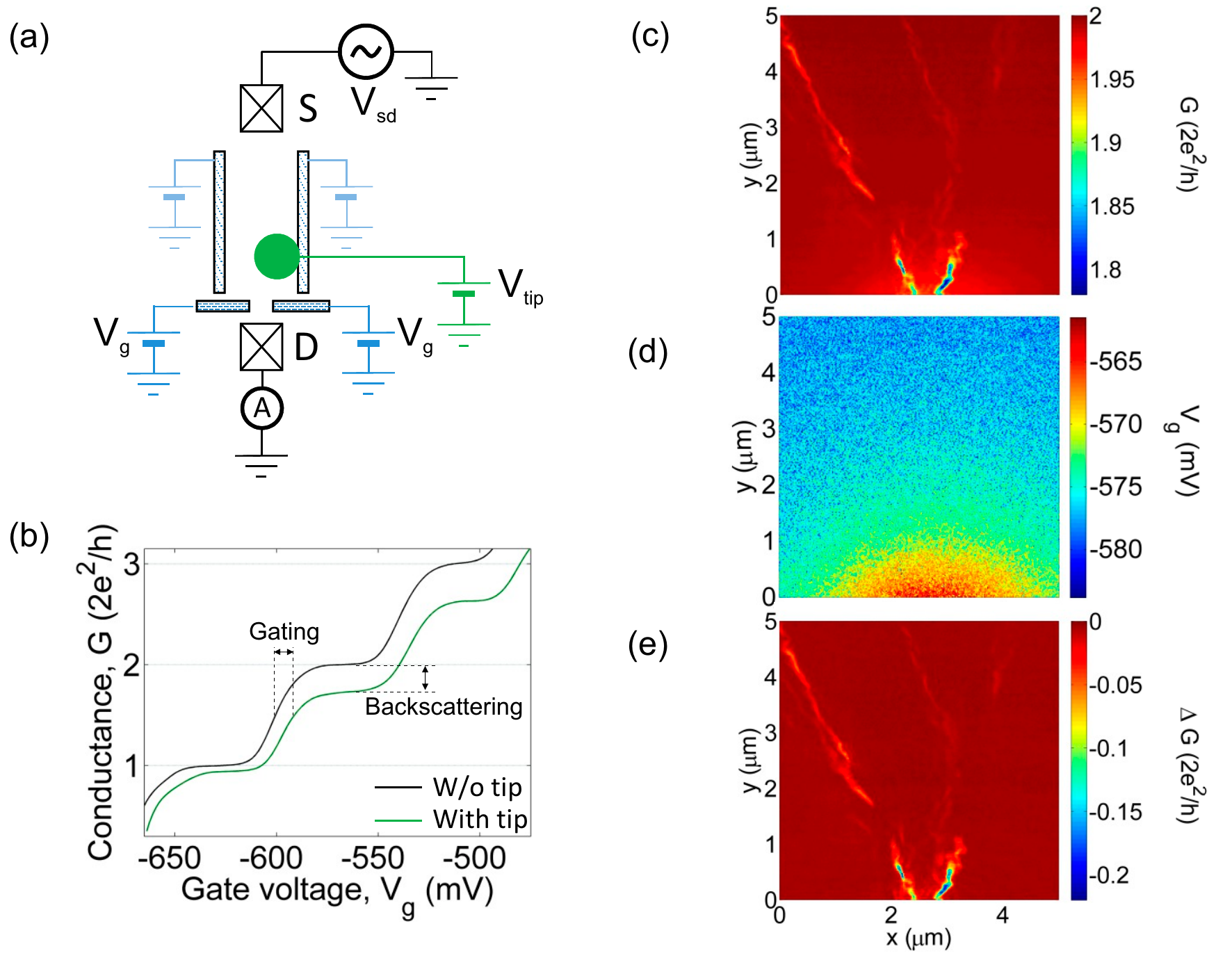}
\caption{(a) Schematics of the setup. A voltage $V_\textrm{sd}$ is applied between source (S) and drain (D). (b) Conductance $G$ as a function of gate voltage $V_\textrm{g}$ with and without the tip. The green curve is obtained when the tip is above a branch. (c) $G(x,y)$ as a function of tip position $(x,y)$. (d) Values of the gate voltage corresponding to the second plateau as a function of tip position. (e) The change of the conductance $\Delta G(x,y)$ in (c) due to electron backscattering when the gating effect is compensated.}
\label{fig:Raw}
\end{center}
\end{figure}

The sample consists of a QPC with a lithographic width of 300 nm and a pair of parallel `channel' top gates separated by 100 nm from the QPC gates, that are 15 $\mu$m long, 150 nm wide and separated by 1 $\mu$m [Fig. \ref{fig:Raw}(a)]. These `channel' top gates are not used in this work, but to compensate strain-related electric fields present below them at zero voltage, we apply a voltage of +200 mV to them. The top gates are Ti/Au electrodes, which are 30 nm high.

Scanning gate measurements are carried out in a He-3 system using a home-built scanning force microscope \cite{Ihn} at 300 mK by applying an ac source-drain voltage of $V_\textrm{sd}=100$ $\mu$V at a frequency of 977 Hz. A metallic tip biased at $V_\textrm{tip}=-6$ V scans at a constant height of 60 nm above the surface inducing a tip-depleted region about 1 $\mu$m in diameter \cite{Steinacher2014a}. The two-terminal current flowing through the sample is converted into voltage using a current-voltage converter with a feedback resistor of 1 M$\Omega$ and a bandwidth of 30 kHz. The conductance $G$ is determined from the current and the applied ac source-drain bias voltage.

\begin{figure}[H]
\begin{center}
\includegraphics[width=16cm]{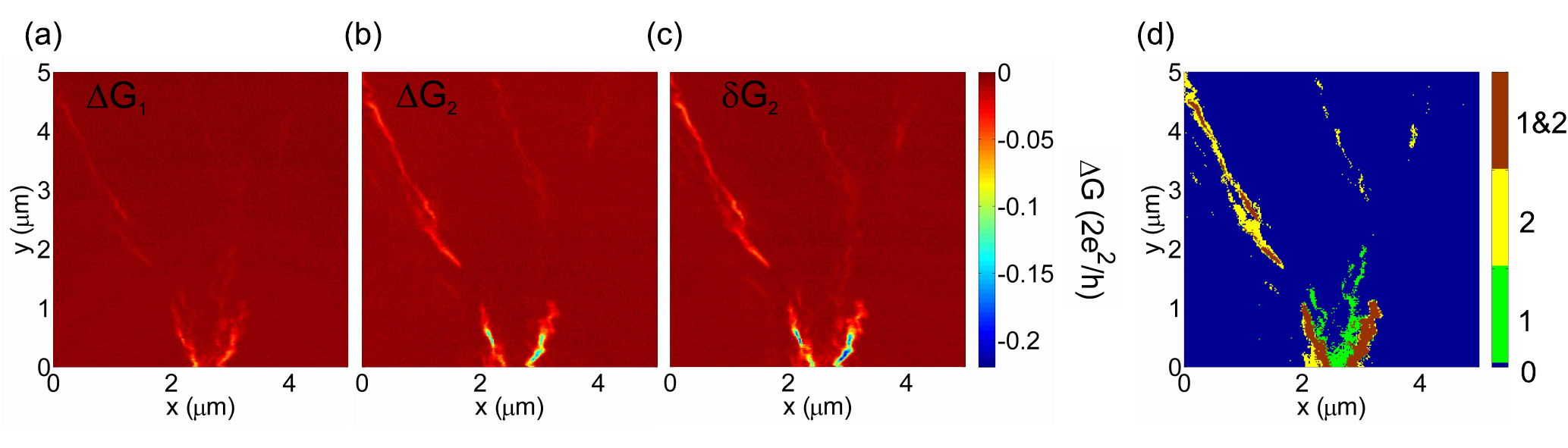}
\caption{Backscattering effect of the tip (a) only on the 1st plateau leading to $\Delta G_1$, (b) only on the 2nd plateau leading to $\Delta G_2$ and (c) on both plateaus without a gating effect. (d) Regions in space (color code) where only the 1st plateau is affected by tip (green), only the 2nd (yellow) and both simultaneously (brown).}
\label{fig:ModeMixing}
\end{center}
\end{figure}

\begin{figure}[H]
\begin{center}
\includegraphics[width=15cm]{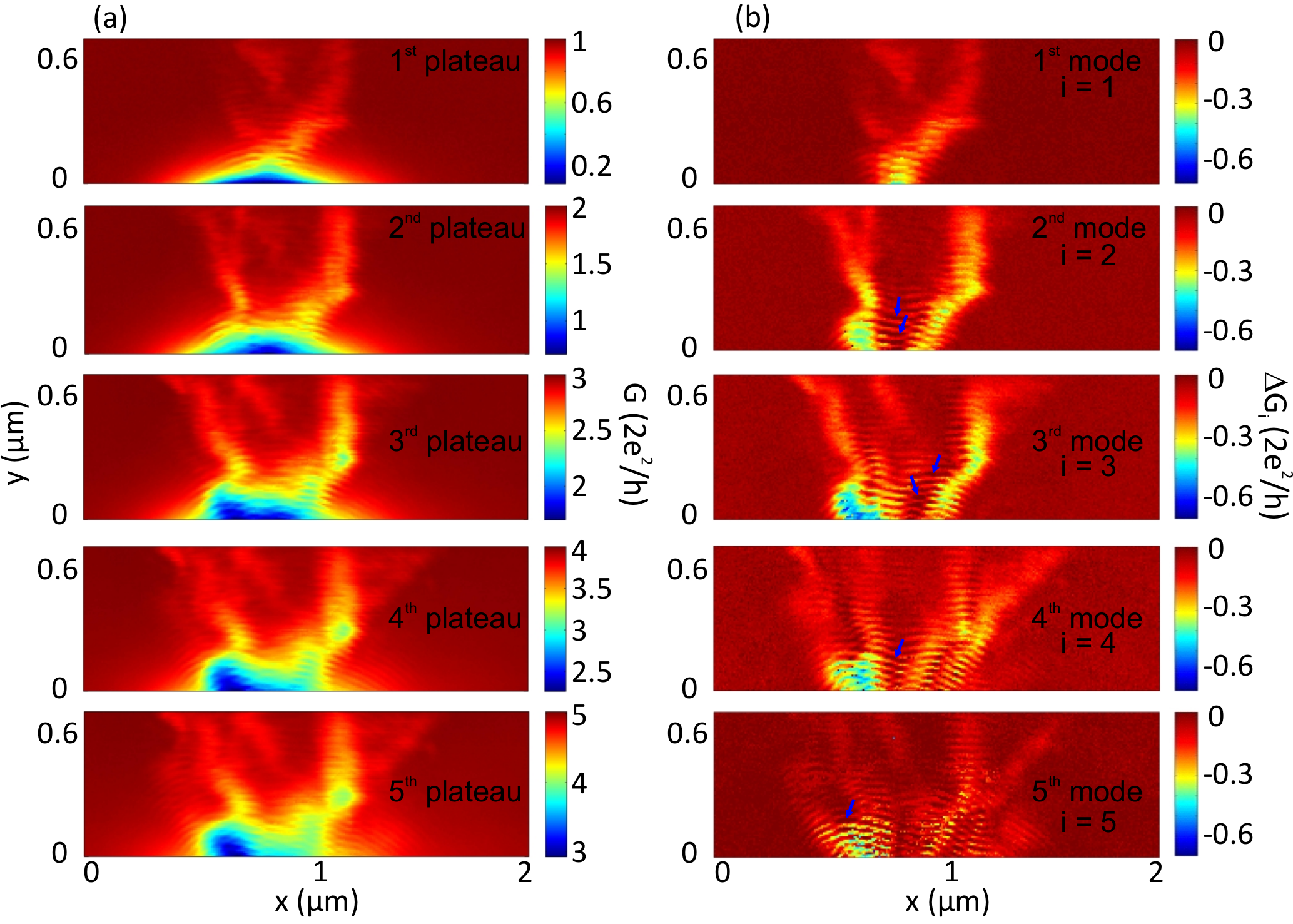}
\caption{(a) Conventional technique: $G(x,y)$ as a function of tip position (raw data) close to the QPC when the constriction is biased on each of the five $G$ plateaus (labelled). (b) Scanning gate grid measurements: $\Delta G(x,y)$ plotted using the technique described in the main text resolving the structure of the first five QPC modes (labelled). Blue arrows mark some of the regions where distorted QPC conductance plateaus manifest themselves in scanning gate maps.}
\label{fig:ModeStructure}
\end{center}
\end{figure}

We carry out two types of measurements. In the first, standard, type, commonly reported in the scanning gate literature, the QPC gate voltage $V_\textrm{g}$ is kept fixed throughout a scan. The resulting conductance maps $G(x,y)$ are plotted as a function of tip position $(x,y)$.
The second type, the investigation and application of which is the focus of this paper, consists in measuring $G(V_\textrm{g})$ as a function of QPC gate voltage at each tip position resulting in $G(x,y,V_\textrm{g})$.
At each tip position a $G(V_\textrm{g})$ curve takes 0.5-1 s depending on the gate voltage range. The pixel separation is 20 nm in Figs. \ref{fig:Raw} and \ref{fig:ModeMixing}, and 8 nm in Fig. \ref{fig:ModeStructure}. The lock-in time constant is 3 ms and the filter slope is 12 dB/oct.

Figure \ref{fig:Raw}(b) compares typical dependences of the conductance through the QPC as a function of top gate voltage with the tip placed about a micron away from the constriction and with the tip withdrawn from the sample surface. Due to electron backscattering off the tip-depleted region the conductance at the plateaus is lowered \cite{TopinkaSci, KozikovBranches} (green curve) below the quantized value (black curve). In 2D conductance maps this deviation from the quantized conductance gives a spatial pattern in which branches and interference fringes can be seen \cite{TopinkaSci, TopinkaNat, Paradiso, KozikovBranches}. In addition, the gating effect caused by the long-range ``tails" of the tip-induced potential capacitively coupled to the QPC shifts $G(V_\textrm{g})$ curves to more positive voltages \cite{KozikovBranches}. The strength of these two effects varies with tip position.

Figure \ref{fig:Raw}(c) shows the conductance as a function of tip position using the standard measurement (first type described above). The QPC is tuned to the second plateau in the absence of the tip and the value of the corresponding gate voltage remains the same throughout the entire scan. Both gating and backscattering effects are clearly seen in this 2D conductance map: branches mark regions of strong backscattering and the gating effect is seen from the color code as a slow decrease of the background conductance from 2 to about $1.95\times2e^2/h$ as the tip moves closer to the QPC. The constriction is about $1~\mu$m away from the lower border of the scan frame, i.e. from $y=0~\mu$m. These observations are in agreement with those of others \cite{TopinkaNat,JuraNatPhys} and with our previous measurements \cite{KozikovBranches}. The gating effect makes it difficult to resolve small variations of the conductance at $0<y<1~\mu$m.

To compensate it with the help of the results of the grid measurement technique, we use $G(x,y,V_\textrm{g})$ curves [Fig. \ref{fig:Raw}(b)] to determine the values of $V_\textrm{g}$ plotted in Fig. \ref{fig:Raw}(d) corresponding to the middle of the second plateau at each tip position $(x,y)$. The gate voltage $V_\textrm{pl}$ at the middle of a plateau corresponds to a zero in the first derivative $dG(V_\textrm{g})/dV_\textrm{g}$, which we determine numerically from the data. We then determine a value of the conductance $G(V_\textrm{pl}(x,y))$ that corresponds to $V_\textrm{pl}$. This procedure performed at each tip position results in compensated conductance maps $G(x,y,V_\textrm{pl}(x,y))$. The result is given in Fig. \ref{fig:Raw}(e) where the value $2e^2/h$ was subtracted. As seen from the uniform background color, the gating effect is compensated and only the backscattering effect (branches) of the tip is present. The shape of the branches at $0<y<1~\mu$m is also uncovered. Subtracting $\Delta G$ in (e) from $G$ in (c) we obtain a smooth background with a local rougness $\Delta G/G$ of less than 0.5\%. In addition, no branches are seen in the $V_\textrm{pl}(x,y)$ map in Fig. \ref{fig:Raw}(d). Fringes originating from interference of backscattered electron waves \cite{TopinkaSci, TopinkaNat, Paradiso, KozikovBranches} are not visible in the images in Fig. \ref{fig:Raw}(c) and (e) due to insufficient spatial resolution intentionally chosen to decrease the measurement time.

The $G(x,y,V_\textrm{pl}(x,y))$ map in Fig. \ref{fig:Raw}(e) has two contributions: a decrease of $G$ for electrons transmitted through the lowest QPC mode, and a decrease of $G$ for electrons transmitted through the second QPC mode.
We plot the effect of the tip on a specific plateau in the following way. We determine the center of plateau $i$ for each point (x,y) using the corresponding minimum at $V_{\textrm{pl},i}(x,y)$ of $dG(x,y,V_\textrm{g})/dV_\textrm{g}$. This gives the conductance $G_i(x,y,V_{\textrm{pl},i}(x,y))$. By subtracting the quantized value $G_{0,i}=i\times2e^2/h$, we obtain $\delta G_i(x,y,V_{\textrm{pl},i}(x,y))=G_i(x,y,V_{\textrm{pl},i}(x,y))-G_{0,i}\leq0$. In addition, by subtracting $\delta G_{i-1}(x,y,V_{\textrm{pl},i-1}(x,y))$ of the previous plateau, we obtain the effect of the tip only on a particular plateau $i$ (we assume that the contribution of plateau $i-1$ to the conductance at the gate voltage of plateau $i$ is the same as at the gate voltage of plateau $i-1$):
\begin{eqnarray}
\Delta G_i\left[x,y,V_{\textrm{pl},i}(x,y)\right]=G_i\left[x,y,V_{\textrm{pl},i}(x,y)\right]-G_{i-1}\left[x,y,V_{\textrm{pl},i-1}(x,y)\right]-2e^2/h
\label{eqn:DeltaG}
\end{eqnarray}

The result of this analysis is shown in Fig. \ref{fig:ModeMixing}(a) and (b). Both plots use the same color scale and the same range on the color axis without any offset. Differences in the intensity and position of branches are clearly seen. For comparison we show $\delta G_2$ in Fig. \ref{fig:ModeMixing}(c) [same as in Fig. \ref{fig:Raw}(e)].

Using (a) and (b) one can plot a map of tip positions at which the effect of the tip on a certain plateau exceeds a given threshold value.  To do this, we introduce a parameter $\beta$ that can take integer values 0 (no branches) and $\beta>0$ (there is a branch in $\Delta G_i$), and choose threshold values $\Delta G_{1,\mathrm{th}}=-0.007\times2e^2/h$ and $\Delta G_{2,\mathrm{th}}=-0.011\times2e^2/h$ in Fig. \ref{fig:ModeMixing}(a) and (b), respectively. The threshold values are chosen by eye to select most of the branches. Then for all values below the threshold of $\Delta G_1$ we set $\beta$=1 and for those below that of $\Delta G_2$ we set $\beta$=2. Then, if $\beta=1,~2$ and 3, the tip affects only the 1st plateau, only the 2nd plateau and both plateaus, respectively.
The result is shown in Fig. \ref{fig:ModeMixing}(d). The branches colored in green are regions where the tip affects only the first plateau, in yellow only the second plateau and in brown both plateaus. The blue color corresponds to areas where there is no backscattering by the tip and therefore $\Delta G_i=\delta G_i=0$. This result shows that as the tip moves across a branch, it can affect several QPC plateaus one after another or simultaneously along its way. At the same time the conductance can exhibit a single dip.

Far from the QPC the main advantage of the grid technique is to separate effects of the tip on each conductance plateau. The gating effect is relatively weak there. However, gating becomes significant when scanning closer to the QPC as seen in conventional SGM images at fixed gate voltages in Fig. \ref{fig:ModeStructure}(a). The border of the scan area at $y=0~\mu$m is about $0.5~\mu$m away from the QPC. As one can see in Fig. \ref{fig:ModeStructure}(a), the tip can reduce $G$ by up to $2\times2e^2/h$ depending on QPC width. As a result, $G$ variations (branches and fringes) are not discernible in the lower half of the images. Nevertheless, interference fringes can already be seen in some regions. Scanning gate grid measurements allow us to compensate the gating effect [Fig. \ref{fig:ModeStructure}(b)] leaving the size of the tip-depleted region to be the only limiting factor preventing to scan even closer to the QPC. As a result, the QPC mode structure and spreading of the interference pattern can be resolved. The technique depends on the width $W$ of the tip-induced potential. The range of $V_\textrm{g}$ needed to compensate the gating effect is larger for larger $W$. In our paper the related change in QPC potential is tiny. As the QPC widens, more modes transmit through it leading to an angular lobe pattern. The number of lobes is equal to the number of modes at the Fermi energy, which in turn is equal to the number of maxima of the squared confined wavefunction $|\Psi|^2$ \cite{TopinkaSci} in the direction normal to the transport axis. For example, for the first QPC mode in Fig. \ref{fig:ModeStructure}(b) one lobe is observed at $0<y<0.15~\mu$m. At $y>0.15~\mu$m it branches out. For the 2nd mode there are two lobes of the angular pattern. For higher modes branching slowly takes over not allowing us to observe the lobe pattern clearly.

In comparison with other works to image angular patterns of individual modes \cite{TopinkaSci}, the presented grid technique allows observing interference fringes coexisting with the angular lobes. It helps compensating the gating effect to study more subtle quantum interference phenomena. This is done point by point in space making it possible to scan large areas neglecting time drifts of the sample as it takes much less time to record the conductance at a single point than in the entire area. We learn how the characteristic QPC curves $G(V_\textrm{g})$ are affected by the tip at each point in space. This information is inaccessible in standard SGM experiments.

\begin{figure}[H]
\begin{center}
\includegraphics[width=15cm]{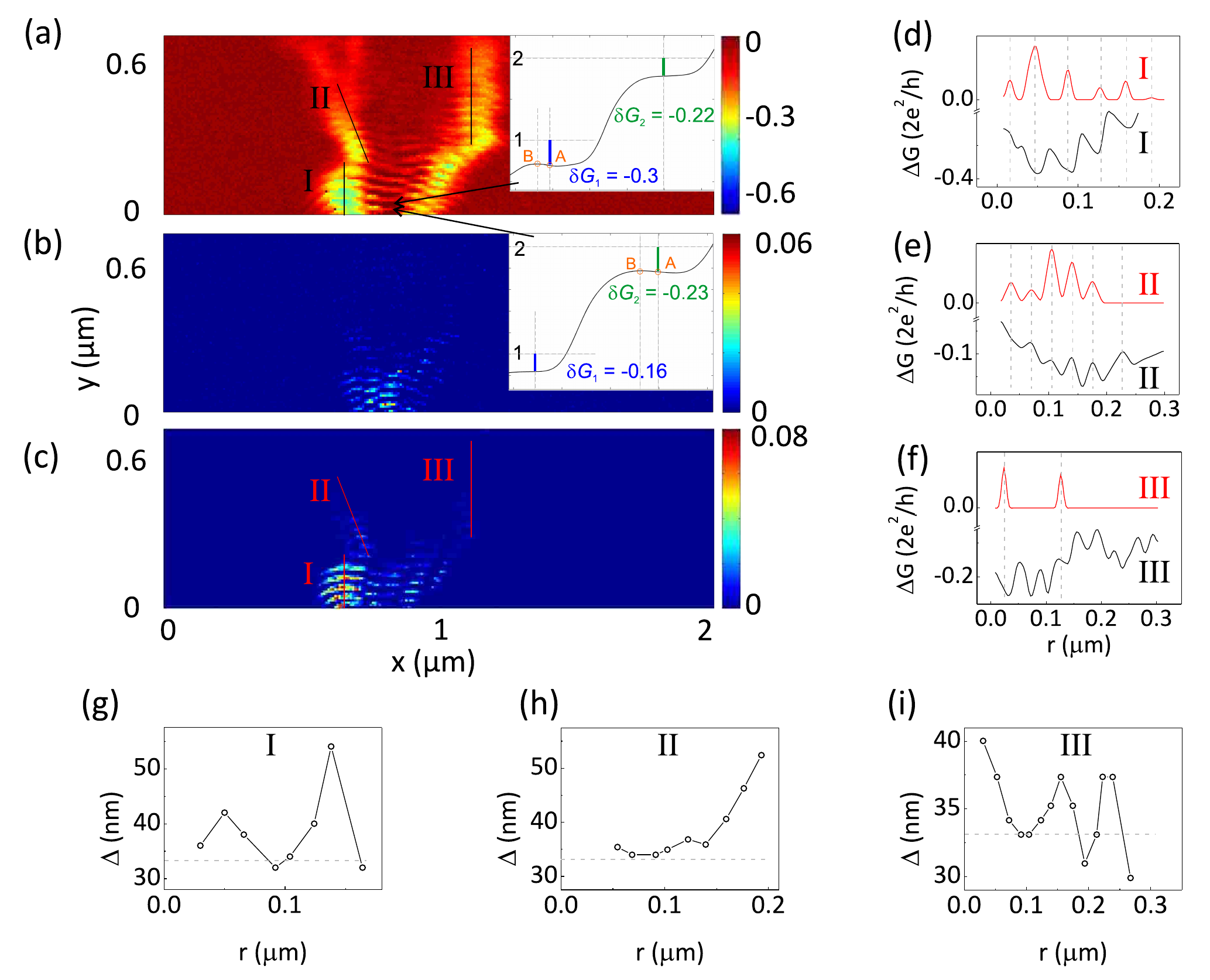}
\caption{(a) $\Delta G(x,y)$ for the 2nd QPC mode. Inset: $G(V_\textrm{g})$ at the tip position marked by the black arrow in the main plot. Dashed lines are guides to the eye. Numbers ``1" and ``2" on the vertical axis correspond to $G$ in units of $2e^2/h$. Short thick vertical blue and green lines indicate deviations of $G$ from 1 and $2\times2e^2/h$, respectively. Values of this deviation in units of $2e^2/h$ are labeled with a corresponding color. Points ``A" and ``B" marked by orange circles correspond to the value of $G$ in the middle between a maximum and a minimum and at a maximum of a distorted plateau, respectively. (b) Difference in the conductance of points ``A" and ``B" at the first $G$ plateau in the inset of (a) in units of $2e^2/h$ as a function of tip position. Inset: see description in (a). (c) Difference $\Delta G_{\textrm{AB,2nd}}$ in the conductance of points ``A" and ``B" at the second $G$ plateau in the inset of (b) in units of $2e^2/h$ as a function of tip position. (d)-(f) Conductance along lines I, II and III in (a) and (c). Colors of the curves correspond to those of the lines in (a) and (c). Vertical lines are guides to the eye. (g)-(i) Fringe spacing along the lines in (a). The horizontal dashed line marks half the Fermi wavelength ($\lambda_\mathrm{F}/2=33$ nm). Zeros along the $r$ axis correspond to the lower ends of the three lines in (a) and (c).}
\label{fig:PositiveG}
\end{center}
\end{figure}

Resolving the QPC mode structure allows us to study it in more detail.
One can notice fringes between the lobes in Fig. \ref{fig:ModeStructure}(b) seen as alternating dark and bright red stripes (some of them are marked by blue arrows as an example). Fringes of the dark red color correspond to $\Delta G_i>0$, $i>1$. [The color scale was limited to $\Delta G_i<0$ in all figures to simplify descriptions of the observed effects. Limiting $\Delta G_i<0$ separates different effects of the tip more transparently.] In Figure  \ref{fig:PositiveG}(a) $G(x,y)$ corresponding to the 2nd mode [as in Fig. \ref{fig:ModeStructure}(b)] is shown. Interference fringes are visible between the two lobes of the angular pattern.

When the tip is placed at a fringe minimum in Fig. \ref{fig:PositiveG}(a) (see upper arrow), the corresponding $G(V_\textrm{g})$ dependence is shown in the inset in Fig. \ref{fig:PositiveG}(a). The first plateau is distorted: there are a local maximum (point ``B" in the inset) and minimum. To resolve the QPC mode structure in Fig. \ref{fig:ModeStructure}(b), $G$ between the maximum and minimum was used, i.e. $G$ at point ``A".
According to Eq. \ref{eqn:DeltaG}, $\Delta G_2=\delta G_2-\delta G_1>0$ as seen from the inset: the deviation of $G$ from $1\times2e^2/h$ is larger than that from $2\times2e^2/h$. For this reason there are tip positions at which $\Delta G_2>0$. We find regions where this occurs to be directly related to regions (tip positions) where the first plateau is distorted. We note that $\delta G_i<0$ always, i.e. the tip never increases $G$ above $i\times2e^2/h$. The difference in $G$ between points ``A" and ``B" is shown in Fig. \ref{fig:PositiveG}(b). For the non-distorted plateau this difference is zero (uniform dark blue color). Regions where it is larger than 0 are tip positions at which the plateau is distorted.
We thus conclude that when a plateau $i$ is distorted, the upper plateau $i+1$ deviates from $(i+1)\times2e^2/h$ less than the distorted plateau $i$ from $i\times2e^2/h$ leading to $\Delta G_{i+1}>0$ for the plateau $i+1$.

A $G(V_\textrm{g})$ curve, which corresponds to a tip position at a fringe maximum in Fig. \ref{fig:PositiveG}(a) (see lower arrow), is plotted in the inset of Fig. \ref{fig:PositiveG}(b). From Eq. \ref{eqn:DeltaG} $\Delta G_2=\delta G_2-\delta G_1<0$ as usually observed in scanning gate experiments.

The interference fringe spacing for different QPC modes is found to be similar to our previous studies \cite{KozikovBranches} in which simultaneous contributions from several modes were studied. Deviations from expected $\lambda_\mathrm{F}/2$ reach $35\%$ in some regions despite the very high quality of our 2DEGs. Using the scanning gate grid technique we relate it to the distorted QPC conductance plateaus.
Indeed, the second plateau in the inset of Fig. \ref{fig:PositiveG}(b) is distorted. We plot in Fig. \ref{fig:PositiveG}(c) the difference $\Delta G_{\textrm{AB,2nd}}$ in $G$ between points ``A" and ``B" on the second plateau as a function of tip position. This difference oscillates in space and in some regions looks similar to the fringes observed in $\Delta G_2$ plotted Fig. \ref{fig:PositiveG}(a). We compare the fringe spacing along lines I, II and III shown in Figs. \ref{fig:PositiveG}(a) and (c). In Fig. \ref{fig:PositiveG}(d)-(f) $\Delta G_{\textrm{AB,2nd}}$ and $G_2$ are plotted along these lines. Maxima of $\Delta G_{\textrm{AB,2nd}}$ in (d)-(f)  (red curve) coincide with minima or maxima of $\Delta G_2$ (black curve) indicating that the fringe spacing in $\Delta G_{\textrm{AB,2nd}}$ and $G_2$ is very similar. The separation between the fringes in $\Delta G_2$ is plotted in Fig. \ref{fig:PositiveG}(g)-(i). We note that deviations of the fringe spacing from $\lambda_\mathrm{F}/2$ for the second QPC mode occur at tip positions at which the second plateau is distorted. For example, lines I and II, along which this deviation reaches about $35\%$, are drawn in areas where the second plateau is distorted [see (c) and the red curve in (d) and (e)]. Along line III the first plateau is sightly distorted only at the beginning of the line [see (f)], which could be the reason for the stronger deviation of the fringe spacing from $\lambda_\mathrm{F}/2$ at the beginning of the curve in (i). In the rest of the curve the fringe spacing deviates from $\lambda_\mathrm{F}/2$ only by about $10\%$.

We therefore propose another scenario for the large fringe separation in addition to those described in our previous work \cite{KozikovBranches}. Regions where distorted plateaus occur can give rise to interference fringes with fringe-spacing significantly deviating from $\lambda_F/2$.
Tip-induced non-adiabaticity of the QPC is likely to be the reason for the distorted plateaus \cite{QPCresonance1, QPCresonance2}. Indeed, the tip-induced potential affects the shape and size of a QPC making the evolution of the wavefunction less adiabatic. This leads to a resonance-like structure in the conductance (distorted plateaus). We rule out the presence of an impurity(s) inside the QPC channel, because measurements of $G$ as a function of asymmetric voltages applied to the two top gates of the QPC \cite{Roessler} revealed only flat or shoulder-like plateaus.

In the analysis used in this paper we assumed that the transmission coefficient for the $i-th$ QPC conductance plateau is the same for plateaus $j>i$. Our finding that the tip distorts plateaus as seen in Fig. \ref{fig:PositiveG}(a)-(c) may indicate that this assumption does not hold for all tip positions. The grid technique allows identifying regions with distorted plateaus and excluding them from the mode-analysis.

In conclusion, we have presented scanning gate grid measurements of a quantum point contact. Its advantages include (i) knowledge of QPC $G(V_\textrm{g})$ curves affected by the tip at each point in the plane, (ii) compensation of the gating effect, (iii) possibility to bring the tip closer to a nanostructure without pinching it off, (iv) imaging and investigating the effect of the tip on a particular QPC mode (plateau) and (v) a single measurement containing $G(x,y)$ in a broad range of $V_\textrm{g}$. With this measurement technique we were able to detect and image branches, across which the tip affects several plateaus at once or one after another on its way, and the mode structure close to the QPC. We also found that the tip can distort conductance plateaus, an effect observable thanks to the grid technique, which indicates tip-induced non-adiabaticity in the QPC. We have argued that it is this effect that leads to the large fringe spacing observed in our scanning gate experiments. We believe that the grid technique can be a powerful tool to gain more information about the studied system at each tip position compared to standard SGM measurements. For example, one can apply the grid technique to phenomena that manifest themselves as shoulders or narrow plateaus observed between regular QPC plateaus at integer multiples of $2e^2/h$, such as the 0.7 anomaly, integer and fractional quantum Hall states, as well as single electron detection. The gating effect will be especially significant here since the QPC conductance is most sensitive to changes in the potential between the regular plateaus.

\begin{acknowledgement}

We acknowledge financial support from the Swiss National Science Foundation and NCCR ``Quantum Science and Technology".

\end{acknowledgement}

Note: The authors declare no competing financial interest.

\bibliography{NewTechnique}

\end{document}